\documentclass[prd,reprint,10pt,nofootinbib]{revtex4-1}
\usepackage{pgfplots}
\usepackage{graphics}      
\usepackage{longtable}     
\usepackage{url}           
\usepackage{bm}            
\usepackage{gensymb}
\usepackage{comment}	   
\usepackage[english]{babel}
\usepackage{physics}
\usepackage{enumitem}
\usepackage{amsmath, amsthm, amsfonts, amssymb, mathtools, calrsfs, tensor, tikz-cd}
\pgfplotsset{width=7cm,compat=1.3}
\usepackage{physics}
\usepackage{xcolor}

\begin{document}
\raggedbottom
\title{Reconsidering astrophysical constraints on macroscopic dark matter}
\author{Jagjit Singh Sidhu, Glenn D. Starkman}
\affiliation    {Physics Department/CERCA/ISO Case Western Reserve University
Cleveland, Ohio 44106-7079, USA}
\date{\today}

\begin{abstract}
Macroscopic dark matter -- ``macros''-- refers to a broad class of alternative candidates
to particle dark matter with still unprobed regions of parameter space.
These candidates would transfer energy primarily through elastic 
scattering with approximately their geometric cross-section. For sufficiently
large cross-sections, the linear
energy deposition could produce observable signals if a macro were to pass 
through compact objects such as white dwarfs or neutron stars
in the form of thermonuclear runaway, leading to
a type IA supernova or superburst respectively.
We update the constraints from white dwarfs.
These are weaker than previously inferred in important respects because of more careful treatment of the passage of a macro through the white dwarf and greater conservatism
regarding the size of the region that must be heated to initiate runaway.
On the other hand,
we place more stringent constraints
on macros at low cross-section, 
using new data from the Montreal White Dwarf Database.
New constraints are inferred 
from the low mass X-ray binary 4U 1820-30, 
in which more than a decade passed 
between successive superbursts.
Updated microlensing constraints are also reported.
\end{abstract}

\maketitle

\section{Introduction}
Dark matter is the most widely accepted explanation for 
cosmological and galactic dynamics (although see \cite{Lelli:2016zqa}), and yet very little is 
known about it beyond some upper limits on
interaction cross-sections over a wide range of masses (see e.g \cite{jacobs2015macro}). 
New fundamental particles, not included in the Standard Model of particle physics, 
are popular candidates because they often arise in models of
Beyond the Standard  Model physics invented for independent reasons (e.g. the axion \cite{axion,Weinberg_axion,Wilczek_axion}).  
However, it remains an open possibility that dark matter is comprised instead 
entirely of macroscopic bound states of fundamental particles. 

Such bound states could avoid strong constraints on the self-interactions of dark matter by virtue of their low number density instead of any intrinsic weakness of their non-gravitational couplings. 
One such open possibility is that dark matter is comprised of bound states of quarks or hadrons, as first proposed by Witten \cite{PhysRevD.30.272} 
as products of a first-order QCD phase transition, and later Lynn, Nelson, and Tetradis \cite{LYNN1990186} and Lynn \cite{1005.2124} again,
who argued in the context of SU(3) chiral perturbation theory that “a bound
state of baryons with a well-defined surface may conceivably form in the presence of kaon condensation.” This would place the dark matter squarely within the Standard Model.
Others have suggested non-Standard Model versions of such nuclear objects and their formation, for example incorporating the axion \cite{hep-ph/0202161}.

Such states are referred to as ``macros.'' 
A macro is then characterized by its geometric cross-section $\sigma_\chi$ and mass $M_\chi$, which are related to the macro's average density $\rho_{\chi}$
\begin{eqnarray}\label{nuclear}
\sigma_{\chi} = 2.4\times10^{-10}{\mathrm {cm}}^2 
	\left(\frac{M_{\chi}}{g}\right)^{2/3}\left(\frac{\rho_\text{nuclear}}{\rho_{\chi}}\right)^{2/3}\,.
\end{eqnarray}
Because of the exciting possibility that macros emerge from essentially the same Standard Model physics as ordinary nuclei, we regard $\rho_\text{nuclear}=3.6\times10^{14}\,$g cm$^{-3}$ as a reference density of particular interest.

Due to their large mass and low number density, macro detectors must be extremely large or experience extremely long integration times to overcome the macros' extremely low flux compared to typical particle dark matter.
Recent comprehensive assessments of limits on such macros as a function of their mass and cross-section \cite{jacobs2015macro,jacobs2015resonant} identify large open windows in 
the parameter space.

For macro
masses $M_x \leq 55\,$g, careful examination of specimens of
old mica for tracks made by passing dark matter \cite{DeRujula:1984axn,Price:1988ge}
has ruled out such objects as the primary dark-matter
candidate (see Figure 1). For $M_x \geq 10^{24}\,$g, a variety of microlensing searches have similarly constrained
macros \cite{Alcock2001,astro-ph/0607207,0912.5297,Griest2013}.
A large region of parameter space was constrained by considering thermonuclear runaways triggered by macros incident on white dwarfs 
\cite{PhysRevD.98.115027}. 
Dark-matter-photon elastic interactions 
were used together with 
Planck cosmic microwave background (CMB) data 
to constrain macros of sufficiently high reduced
cross-section $\sigma_x/M_x$ \cite{1309.7588}.
Prior work had already constrained 
a similar range of parameter space, 
by showing that 
the consequence of dark-matter interactions 
with Standard Model particles 
is to dampen the primordial matter fluctuations 
and essentially erase all structures 
below a given scale (see e.g. \cite{Bhm}).
The region of parameter space 
where macros would have produced a devastating
injury similar to a gunshot wound 
on the carefully monitored population 
of the western world 
was also recently constrained \cite{1907.06674}.

Recently, together with collaborators, 
we suggested how ultra-high-energy cosmic-ray detectors 
that exploit atmospheric fluoresence 
could potentially be modified 
to probe parts of macro parameter space \cite{Sidhu:2018auv}, 
including macros of nuclear density and intermediate mass.
This analysis has led to constraints being placed
using networks of cameras 
that were originally built to study bolides, 
i.e. extremely bright meteorites 
with absolute magnitude $M_v \leq -5$ \cite{1908.00557}. 
We have also suggested how the approach 
applied to mica \cite{DeRujula:1984axn,Price:1988ge} 
could be extended to a larger, widely available sample 
of appropriate rock \cite{1905.10025}, 
and used to search for larger-mass macros. 

There remains a wide range of parameters 
$M_X$ and $\sigma_x$ 
that are currently unconstrained 
by all the previously mentioned works.
This includes a wide range 
of the nuclear-density line \eqref{nuclear}.

In this paper, 
we update certain astrophysical constraints 
on 
macro-dark-matter mass and cross-section.
As in previous works, 
we consider dark matter of a single mass,
even though a broad mass distribution 
is a reasonable possibility 
in the context of a composite dark-matter candidate. 

The rest of this paper is organized as follows. 
In section \ref{sec:lensing}, 
we update constraints on macros 
from microlensing and femtolensing. 
In section III, 
we update constraints from the observation
(or not) of Type Ia supernova events in
white dwarfs, 
weakening previous contraints by more careful consideration
of the propagation of the macro through the white dwarf,
and through greater conservatism regarding the minimum 
size region that must be heated above a critical temperature.
On the other hand, we show that the large number of 
white dwarfs that have been collected from the literature
in the Montreal White Dwarf Database \cite{MWDD} 
allow more stringent
constraints to be place on macros at low cross-sections.
We also produce new constraints using arguments about thermonuclear runaway, by
applying them to neutron stars, 
and the observation of superbursts. 
In the process, 
we rule out macros of ``nuclear density'' 
as the sole dark-matter candidate 
for certain mass ranges.

\section{Lensing Constraints}
\label{sec:lensing}

Massive objects passing between a light source 
and the Earth will gravitationally lens the source. 
For appropriately located objects of sufficient mass, 
the image can be measurably affected -- 
it can be amplified, distorted, 
or even split into multiple images some of which are amplified and others de-amplified 
compared to the unlensed source.  
These effects are time-dependent, 
and can change over observationally accessible time-frames 
as the source-lens-observer geometry changes
with the relative motion of elements.

\subsection{Microlensing of M31}
A recent seven-hour high-cadence observation 
of M31 (Andromeda),
using the Subaru Hyper Suprime-Cam (HSC),
was used to derive bounds 
for primordial black holes (PBHs) \cite{Niikura2019}, 
based on the non-observation of a ``lensing event'' -- 
the time-dependent amplification of a source star. 
The amount of dark matter 
that would be expected to pass through
the ``lensing tube'' \cite{Griest1991} 
during this observation can readily be calculated. 
For dark-matter candidates of a particular mass,
the lensing tube represents 
the volume along the line-of-sight
where the presence of a lens 
would cause a lensing event with amplification and duration
greater than the appropriate threshold values for detectability. 

Null lensing results allow constraints 
to be placed on the abundance of dark matter objects 
within a certain mass range,
given the sensitivity  of the measuring telescope.
Objects that are too large or too small in mass
would not be expected to produce observable lensing events. 
Objects of too-low mass produce 
lensing events with a low amplification.
Morover, the duration of a lensing event
is approximately the time it takes for the lens 
to cross its Einstein radius \cite{Niikura2019}
\begin{equation}\label{Einsteinradius}
	R_E =\sqrt{\frac{4GM_{\chi}}{\pi c^2}d\left(1-\frac{d}{d_s}\right)}\,,
\end{equation}
where $d$ is the distance to the macro 
and $d_s \approx 770$kpc is the distance to M31. 
We assume for simplicity that the lens is in the Milky Way,
so $d \ll d_s$. 
The crossing time
\begin{align}
t_E &\equiv \frac{R_E}{v}
\approx 34 min \left[\frac{M_{PBH}}{10^{-8}M_\odot}
\frac{d}{100kpc}
\frac{v}{200 km/s}\right]^{\frac{1}{2}}\,. 
\end{align}
Lensing events by low-mass objects are therefore brief,
and likely to be missed between successive observations.

For a relatively short observational ``campaign,''
objects that are too massive produce lensing events 
that are too long 
for the change in brightness of the lensed star 
to be detected. 
For the seven-hour long observation of M31 using
Subaru Hyper Suprime-Cam (HSC),
the maximum sensitivity to lensing events was for events 
with $t_{FWHM} \approx [0.07,3]\,$hours.
With the mass density of dark matter 
fixed by galactic dynamics,
the number of lensing events is also impacted
by the  number density of candidate dark matter lenses
being inversely proportional to their mass.
 
Reference \cite{Niikura2019} has shown that 
diffraction effects become important 
and the maximum image magnification 
is significantly reduced for
$M_{PBH} \lesssim 10^{23}\,$g, 
as the Schwarzchild radius of the PBH 
becomes smaller than the photon wavelength\footnote{
	This does not mean that the Schwarzschild radius, $R_s$, 
	can be interpreted as the effective radius of the lens --
	the typical impact parameter of photons involved in the 
	lensing event is still $R_E$ rather than $R_s$.},
with no constraints 
for $M_{PBH} \leq 10^{22}\,$g.

Since  the microlensing interaction 
is purely gravitational, 
these results \cite{Niikura2019} 
are directly applicable to compact objects other than PBHs,
such as macros. 
We may therefore rule out macros with 
$10^{22}\mathrm{g} \leq M_\chi \leq 4\times10^{24}\mathrm{g}$ as the dominant dark matter component.
Above $M_\chi \leq 4\times10^{24}$,
they have already been excluded 
by previous microlensing experiments
(see \cite{jacobs2015macro} and references therein). 



The microlensing constraint does not apply 
if the macro  blocks a significant fraction 
of the light that would be amplified. 
This happens when the macro radius exceeds the 
Einstein radius.
The microlensing bound therefore applies only to
\begin{equation}
	\sigma_{\chi} \leq \frac{4GM_{\chi}}{c^2}d 
	= 7\times 10^{-4}\frac{M_{\chi} }{\text{g}} \text{cm}^2.
\end{equation}
However, this value of $\sigma_{\chi}$ 
lies well within the region of parameter space
already ruled-out 
by consideration of interactions 
between macros and CMB photons 
(shaded grey in Figure 1). 

\subsection{Femtolensing}

Femtolensing refers to gravitational lensing where the angular separation between two lensed images of the same source is of order $10^{-15}$ arcseconds. 
The separate images cannot be resolved, however an interference pattern in the energy spectrum of background sources could be observable.

In \cite{1204.2056}, constraints were placed on the abundances of PBH with $10^{17}\,$g $\lesssim M_{PBH} \lesssim 10^{20}\,$g, from the non-observation of such interference patterns in gamma-ray bursts (GRBs).
However, \cite{1807.11495} revisited these constraints,
 taking into account  the finite size of the GRB sources,
  among several additional corrections. 
As the emission size of the GRB grows, 
changes in the magnification spectrum 
are damped more strongly, 
until they eventually disappear once the emission size
exceeds the  Einstein radius.
For realistic emission sizes of gamma ray sources --
$a_s \geq 10^{10}$cm -- 
the femtolensing constraint is removed entirely
(see e.g. Figure 2 of \cite{1807.11495}).

The discovery of a large number of sources 
with $a_s \leq 10^9$cm 
would allow limits to be placed \cite{1807.11495} 
on the abundance of dark matter in compact objects 
with $10^{16} \leq M_{CO} \leq 10^{19}$g. 
The smaller the sources,  
the more stringent the constraints 
(see Figure 5 of \cite{1807.11495}).

\section{Constraints from thermonuclear runaways}
\label{sec:thermonuclear}

Reference \cite{Timmes1992} showed that
in a white dwarf 
a sufficiently large localized injection of energy 
might trigger thermonuclear reactions. 
If a critical temperature were exceeded 
in a region of sufficient size, 
fusion of carbon atoms 
would initiate subsequent reactions 
before the heat was able to diffuse away. 
This chain reaction may lead to thermonuclear runaway,
and the white dwarf
would undergo 
a type IA supernova explosion.
The minimum temperature needed to trigger such reactions 
is  \cite{Timmes1992} $T_{trig}\sim 3\times 10^9\,$K. 
The minimum size of the trigger region $\lambda_{trig}$
depends on the local density. 
Neutron stars can also exhibit thermonuclear runaway in 
the form of a superburst \cite{1702.04899}.
Thus, it may be that a similar mechanism could
trigger a superburst in neutron stars.

Using the analysis of \cite{Timmes1992}, 
the authors of \cite{1505.04444} 
placed constraints on the abundance of PBHs 
in the mass range  
$10^{19} \lesssim M_\text{PBH} \lesssim 10^{20}$g 
from the continued existence of old white dwarfs.
If a PBH traveled through a white dwarf, 
the adjacent matter 
would be gravitationally accelerated toward its trajectory. 
Upon thermalization, the temperature would exceed $T_{trig}$. 
If the PBH was sufficiently massive, 
the size of this heated region would exceed $\lambda_{trig}$.
Constraints were therefore placed 
on the abundance of PBHs of that or greater mass. 
However, it was subsequently shown
\cite{MonteroCamacho2019} that the order-of-magnitude
estimates employed by Graham {\it et al.} \cite{1505.04444}
likely constrained PBHs with masses that were too small to cause thermonuclear runaways. It is unlikely that
thermonuclear runaway can constrain any significant 
portion of the PBH parameter space below $M_{PBH} \sim 10^{22}\,$g.

Similar constraints were obtained \cite{PhysRevD.98.115027} 
on dark matter that deposits energy in the white dwarf 
by elastic scattering --
objects that we term macros.
Here the region of the white dwarf that is heated above
$T_{trig}$ is potentially much larger, and so the
limits are likely more robust.
We will proceed with the formalism 
developed in \cite{1505.04444,PhysRevD.98.115027} 
to re-examine the regions of macro parameter space
in which a macro would have produced 
observable consequences for white dwarfs or neutron stars. 
We first constrain macros 
that would have initiated a superburst on a neutron star 
in a shorter time than that which occurs naturally.
Next,
we show the upper bound derived in \cite{PhysRevD.98.115027}
for macros triggering type IA supernovae in a white dwarf
is too stringent and derive a more accurate upper bound.

\subsection{Size of the trigger region}\label{sec:trig}

In \cite{Timmes1992} $\lambda_{trig}$ was calculated for  
$5\times 10^7$g cm$^{-3} \leq \rho 
	\leq 5\times 10^9$g cm$^{-3}$; 
however, the outer regions of white dwarfs 
have densities as low as $10^7$g cm$^{-3}$ 
and lower-mass white dwarfs are nowhere denser 
than approximately $10^7$g cm$^{-3}$.
To extend the analysis to lower density, 
we use the scaling relation 
between $\lambda_\text{trig}$ 
and the local density \cite{1505.04444},
obtained by comparing the heat diffusion rate 
to the carbon fusion rate.
Timmes and Woosley showed \cite{Timmes1992} 
that the resulting estimate of $\lambda_{trig}$ 
was within a few per cent of the value obtained 
by simultaneously solving the hydrodynamics, 
nuclear kinetics  and transport equations 
of the deflagration front 
as it propagates through the white dwarf. 

Nevertheless, it is still unclear how exactly type IA supernovae
are initiated \cite{Rpke2017}. Thus, while the initiation of a deflagration flame
may be a necessary condition for a type IA supernova, it may not
be sufficient. We will proceed with the caveat that for
the constraints produced in this section it must still be shown
through simulations that those regions of macro parameter space 
can indeed produce catastrophic thermonuclear runaway.
The situation is similar with neutron stars where the precise
nature of the initiation of a superburst is unclear \cite{1702.04899}
and so in this case as well we also proceed with the understanding
that the regions of macro parameter space constrained in this paper
still need to be tested against simulations.

The characteristic time for heat to diffuse 
through (and out of) a region of size $\lambda_{trig}$ is
\begin{equation}\label{Heateqnchartime}
	\tau_{{diffusion}} \approx \frac{\lambda_\text{trig}^2}{\alpha}\,,
\end{equation}
where $\alpha =K/c_p \rho$ 
is the thermal diffusivity of the medium, 
$c_p$ is the specific heat capacity,
$\rho$ is the density,
$K\propto T^3/(\kappa \rho)$ 
is the thermal conductivity\cite{9783540502111}, 
and $\kappa$ is the opacity 
of the dominant carrier of energy. 
For $\rho<10^8 \text{g cm}^{-3}$, 
photons are the dominant energy carriers.
As most of the electrons are ionized, 
free-free transitions are the main source of opacity, 
and so \cite{9783540502111} $\kappa \propto \rho$. 
 
The mean free time between carbon-ion collisions 
(and consequently fusion reactions) 
is inversely proportional to the density,
\begin{equation}
\label{fusiontime}
\tau_{fusion} \sim \left(\sigma v n\right)^{-1}.
\end{equation} 
Since the size of the trigger region 
is obtained by requiring $\tau_{fusion}<\tau_{diffusion}$,
$\lambda_\text{trig} \propto \rho^{-2}$. 
  
The presence of iron-group impurities 
in the heavy-element ocean 
located near the surface of a neutron star
reduces the number density of carbon ions, 
This increases the minimum column density 
that must be accreted  
for thermonuclear runaway to be achieved \cite{Strohmayer2002}.

\subsection{Energy deposition}
As a macro passes through a compact object,
it causes an initial rise 
in the temperature of the affected matter -- 
from $T_i$ to $T_f$ -- through elastic scattering.
The magnitude of this temperature rise 
depends on the energy deposited 
per unit mass of white-dwarf or neutron-star material
\begin{equation}
\label{eq:specificheat}
\varepsilon_{in} = \int_{T_i}^{T_f}C_V dT\,,
\end{equation}
where $C_V$ is the specific heat.
The usual temperature of a white dwarf or neutron star 
is at most only $T_i\sim 1\,$keV. 
This is much less than $T_{crit}$, and so can be neglected.

The ions in a white dwarf are non-degenerate and have 
specific heat \cite{balberg2000properties}
\begin{equation}\label{T1}
	C_{V,ions} = 3 \frac{k_B}{\mu}\,,
\end{equation}
where $\mu \approx 1.7 m_p$ is the mean molecular weight 
and $m_p$ is the proton mass.

For large densities, $\rho\gtrsim 10^9 {g cm}^{-3}$, 
the Fermi energy approaches the thermal energy and 
$C_V$ acquires a contribution 
from a relativistic degenerate 
electron gas \cite{astro-ph/0107213}
\begin{align}\label{T2}
	C_{V,electrons} = 
	\left(\frac{\pi^2 Z k_B}{A m_p}\right)\left(\frac{k_B T}
	{E_{F}}\right)\,.
\end{align}
Here $Z$ is the average atomic number of the material,
and $A$ is the average nucleon number.
The Fermi energy of the degenerate electron gas 
\begin{align}
	E_{F} = 
	1.9{MeV}\left(\frac{2Z}{A}\right)^{\frac{1}{3}}
	\rho_8^{\frac{1}{3}},
\end{align}
with $\rho_8$ the density 
of that region of the compact object 
in units of $10^8$g cm$^{-3}$. 
$E_F$ is typically a few MeV 
for the range of densities 
where the degenerate-electron specific heat 
contributes significantly.

Using \eqref{T1} and \eqref{T2} 
in \eqref{eq:specificheat},
\begin{align}
\frac{3}{2}\frac{k_B T}{\mu} + \left(\frac{\pi^2 Z}{2A }\right)\frac{(k_B T)^2}{m_p E_{F}} = \varepsilon_{in}(v_x, \sigma_x, M_x)
\end{align}
The first term on the left hand side  dominates 
for $k_BT \lessapprox 1MeV$, 
above which the second term dominates.

$\varepsilon_{in}$, and hence $T_f$, 
could depend on the velocity of the macro $v_{\chi}$, 
e.g. in the case of elastic scattering. 
A macro that is incident on a white dwarf or neutron star
will have been gravitationally accelerated 
to much greater than its initial speed, thus
\begin{equation}\label{initialspeed}
v_\chi \simeq v_{esc} 
	\simeq \sqrt{\frac{2GM_{CO}}{r_{CO}}} \,,
\end{equation}
where $M_{CO}$ is the mass of the compact object
and $r_{CO}$ is the radius of the compact object.
We have neglected  relativistic corrections because, 
even for neutron stars, they are only 3$\%$.
If the final temperature of the heated region
$T \geq T_{crit} \sim 10^{10}\,$K, we expect some of this energy to
be used in endothermic photodisintegration reactions. Thus,
the size of the region with $T>T_{trig}$ is not
expected to be much larger than $\sigma_x$.
For example, at $T = 10^{10}\,$K,
the carbon photodisintegration rate to three alpha
particles (the reverse of the triple-alpha process)
is $\sim 10^{10}\,$s$^{-1}$.
The carbon fusion rate \cite{Caughlan1988} 
at this temperature and $\rho = 10^7\,$g cm$^{-3}$
is $\sim$ few $\times 10^{10}\,$s$^{-1}$.
However, due to the strong temperature dependence
of these rates, 
fusion becomes the fastest process below $T \sim 10^{10}\,$K.
We therefore expect the propagation of a deflagration flame
to begin once the temperature drops below $T \sim 10^{10}\,$K

In the end, we require
\begin{equation}\label{sigmamin}
	\sigma_x > 10\frac{\pi}{4} \lambda_{trig}^2\,,	
\end{equation}
and that enough energy is deposited to 
raise that area $\sigma_x$ above $T_{trig}$:
\begin{equation}
	v_x > \sqrt{C_V T_{trig}}\,.
\end{equation}
The factor of 10 in \eqref{sigmamin} is to ensure that 
photodisintegration does not quench the deflagration flame -- 
by requiring a region much larger than the trigger size to be heated above $T_{trig}$, 
we expect the propagation of a deflagration flame
to be more likely.

\subsection{Elastic scattering bounds}

We consider the bounds on the 
$\sigma_x - M_x$ parameter space from
energy deposition by the macro into the 
compact-object through elastic scattering.
We begin by showing that the maximum value of $\sigma_x$ 
for the constrained region, 
previously obtained in \cite{PhysRevD.98.115027}
through the study of a population of white dwarfs,
was overly optimistic (i.e. too high);
we derive a more accurate (lower) value.
We also derive the analogous quantity for neutron stars,
subject to some additional assumptions
due to the more extreme nature of the environment
compared to white dwarfs.

Next, we determine the lower boundary of the 
constrained region 
(i.e. the smallest $\sigma_x$ for each $M_x$),
which is the minimum cross-section
necessary to initiate the propagation of a deflagration
flame in a white dwarf or on the surface of a neutron star.
Finally, we determine what mass ranges can be probed
by each of white dwarfs and neutron stars, 
to determine the constraints 
subject to the caveats described above.

Both white dwarfs and neutron stars
that are in binary systems
can undergo thermonuclear runaway.
White dwarfs (WDs) undergo a type IA 
supernova event if the WD
accretes enough mass from its
companion for its mass to reach the 
Chandrasekhar limit, the maximum 
possible white  dwarf mass
that  can  be  supported  by electron  
degeneracy  pressure \cite{Chandra}.

Unlike WDs, 
NSs will not explode catastrophically. 
The  outer layer of a NS consists of 
an ocean of heavy elements  
including a significant amount of carbon 
at high  densities 
($\mathcal{O}(100$ m$)$ below the surface). 
Ignition of this carbon layer 
can cause a NS to undergo a ``superburst'' (\cite{2011ATel}. 
These have been observed with recurrence times
ranging from a few days to $\sim10$ years \cite{1702.04899}.
Typically, 
superbursts occur once the mass 
of the layer of carbon -- 
formed from the accretion of hydrogen or helium onto the NS --
reaches $\sim 10^{24}$g \cite{Strohmayer2002,Cumming2001}. 
Heat flowing out from the crust 
is deposited in the carbon ``ocean''
by electron capture and pycnonuclear reactions
\cite{Cumming2001},
augmenting compressional heating by the overlying material.
Only once this much material has accumulated 
is the base of the carbon ocean 
hot enough 
to initiate a thermonuclear runaway.
This yields a superburst energy of $\sim 10^{42}$ergs,
assuming all the carbon ignites.

\subsubsection{Maximum constrained reduced cross-section}
The energy deposited through elastic scattering
by a macro transiting a compact object is
\begin{equation}
\frac{dE}{dx} = \sigma_x \rho_{CO} v_x^2\,,
\end{equation}
where $\rho_{CO}$ is the local density at a point
within the compact object.
$v_x$ is a function of the depth of the macro in the compact object. 
The drag force experienced by a macro
will decelerate it  
\begin{equation}\label{acc}
a_x = \frac{GM_{CO}}{R_{CO}^2} -
\frac{1}{2} C_d \rho_{CO} v_x^2 \sigma_x/M_x\,.
\end{equation}
Here $C_d$ is the drag coefficent, 
and depends on the Reynolds number,
$Re\,$, of the flow
\begin{equation}
Re = \frac{\rho u L}{\mu}\,,
\end{equation}
where $u = v_x$ is the relative velocity between the macro
and the material of the compact object, $L = r_x$ is the characteristic
length scale of the problem, and $\mu$ is the dynamical
viscosity, which is not known for white dwarfs or the outer regions
of a neutron star. 
For values suggested from theory \cite{DallOsso2014}, 
the range of $Re$ 
for our purposes here
is never low enough for the drag coefficient to deviate from
the typical value 
for a sphere of $0.1 - 2$ \cite{Duan2015}. 
However,
for low values of $Re$, $C_d$ could increase by several
orders-of-magnitude, resulting in the macro experiencing
a much higher drag force. 
If the dynamical viscosity is subsequently
determined to be significantly higher, 
this would further reduce the parameter space that may be probed by thermonclear runaway.
For now, we proceed with the most conservative value in the standard range, $\mathrm{C_d}=2$,
in producing our constraints.

To solve \eqref{acc}, which we do numerically,
we must use a 
density profile for a typical compact object.
We use the density profile of a typical
white dwarf \cite{timmes} and of a typical
neutron star crust\cite{Datta1995} to
simulate the evolution
of the velocity of an incident macro
for various values of
$\sigma_x/M_x$. For macros
with a sufficiently high
$\sigma_x/M_x$, the macro
is slowed down before it reaches the
the relevant depth in a 
compact object
and is unable to transfer
enough energy to trigger thermonuclear runaway.
We find this limiting value of the
reduced cross section to be 
$\sigma_x/M_x \gtrapprox 10^{-16}$ 
cm$^2$g$^{-1}$ for white dwarfs and
$\sigma_x/M_x \gtrapprox 10^{-12}$ 
cm$^2$g$^{-1}$ for neutron stars.
However, there is a narrow range of values of $\sigma_x/M_x$
around these two values where sufficient energy
is transferred to initiate thermonuclear runaway
that is dependent on where exactly in the compact object
thermonuclear runaway is initiated.

The upper bound on the reduced cross-section
for white dwarfs is significantly smaller than
that derived in \cite{PhysRevD.98.115027}.
In reference \cite{PhysRevD.98.115027}, the macro
was assumed to be able to trigger a type IA supernova
once it penetrated the non-degenerate surface layer of
a white dwarf, which is typically
narrow and much less dense than central densities. 
This assumption overestimated
the parameter space that was constrained. For a given cross-section, $\sigma_x$,
macros of too small a mass $M_x$ were constrained. 
In this work, we have used a typical white dwarf density profile
from \cite{timmes} to better estimate the true boundary
from white dwarfs. 
This is itself  uncertain,
since the radial density profile of white dwarfs
has not been determined definitively -- 
the correct bound could lie above or below our bound, however, the upper bound in 
\cite{PhysRevD.98.115027} is indeed an overestimate.

We find that neutron stars might ``re-constrain''
some of the parameter space 
that was previously ruled out by white dwarfs.
However, these neutron-star constraints 
merit additional scrutiny 
due to the relativistic speeds 
reached by macros incident on the surface of a neutron star,
$v \sim 0.7$c.
We require that the macro not be destroyed in transiting 
the outer layers of the neutron star 
before reaching the heavy-element ocean. 
The exact constrained region therefore depends 
on the microphysics of the macro,
and how tightly it is bound.
We can get an estimate of the constraints
by taking the macro to be made of baryons,
and estimating that 
the logarithm of the binding energy per baryon $E_b$
scales linearly with the logarithm of the density, 
between atomic density 
	($\rho_{atomic}\simeq 1g/cm^{-3}$,
	 $E_b\simeq 10$eV)
and nuclear density 
	($\rho_{nuclear}\simeq 10^14g/cm^{-3}$,
	 $E_b\simeq 1$MeV).
This yields an expression 
for the scaling between binding energy and density
\begin{equation}
E_b \sim 10 eV 
	\left(\frac{\rho_x}{g/cm^{-3}}\right)^{\frac{3}{7}}\,,
\end{equation}
where of course 
\begin{equation}
\rho_x = \frac{3 M_x \pi^{\frac{1}{2}}}{4 \sigma_x^{\frac{3}{2}}}\,.\nonumber
\end{equation}
Crudely, we require the energy transferred to be less than
the binding energy per baryon multiplied by the
number of baryons in the macro
\begin{equation}
E_b \frac{M_x}{m_b} \geq \rho \sigma_x v_x^2 L\,.
\end{equation}
where $m_b < 940 MeV$ is the mass of a baryon. 
(This ignores the very definite possibility 
of ablation of the macro surface.)
This enforces
a bound similar to that found above, 
$\sigma_x/M_x \lesssim 10^{-11}$ 
cm$^2$g$^{-1}$\.
The exact constrained region depends on the microphysics
of the macro and the details of how it is held together.
However, the upper limit in $\sigma_x$ on the constrained region comes from considering the drag on the macro
through the overlying layers of the compact object,
which is more stringent than the considerations 
of binding energy.

\subsubsection{Minimum cross-section of constraint region}

For the elastic-scattering mechanism, 
$\varepsilon_{in} = v_C^2$, 
i.e. whether or not $E_{trig}$ is reached 
depends -- as discussed earlier -- 
on the speed of the macro as it impacts the carbon atoms. 

The lower bound on $\sigma_x$ 
was determined using \eqref{sigmamin}.
 Since $\lambda_{trig}\propto\rho^{-2}$,
 it varies 
 along the trajectory of the macro
 through the white dwarf or neutron star crust.
 For smaller values of $\sigma_x/M_x$ the macro will
 deccelerate to the minimum speed at which it can still
 trigger thermonuclear runaway at a greater depth,
 corresponding to a higher density,
 and consequently a smaller $\lambda_{trig}$.
Given \eqref{sigmamin},
for smaller values of $\sigma_x/M_x$,
smaller values of $\sigma_x$ can be probed.
Requiring that the macro not lose appreciable kinetic energy
through the non-degenerate surface layer of a white dwarf is
not a sufficient requirement for triggering thermonuclear runaway.
For white dwarfs, the trigger sizes are given

in \cite{Timmes1992}.
For neutron stars, $\lambda_{trig}$ is now larger because the heavy element ocean in a neutron star is expected to be only $\sim 20\%$ carbon \cite{Strohmayer2002}. Thus, the mean free time between collisions of carbon atoms increases since the number density of carbon atoms decreases. The diffusion rate decreases because the number 
density decreases, although the specific heat and thermal conductivity aren't changed significantly. Thus, $\lambda_{trig}$ increases. 

For neutron stars, 
to obtain the most restrictive constraints,
we take into consideration
the evolution
of the column density as matter
is accreted from the companion in the binary system. 
As accretion proceeds, 
the underlying layers are compressed to higher densities. 
Thus, some time is required 
to form carbon of a certain minimum density. 
Although the maximum effective 
exposure time for the neutron star
we use to place constraints is $T = 2.5\,$yr
(as discussed below),
the denser the carbon, 
the shorter the period for which it is ``exposed''.
Consequently, 
smaller cross-sections (corresponding to
smaller $\lambda_{trig}$ and higher densities) 
can only be probed for smaller-mass macros, 
which have higher fluxes.



For white-dwarf constraints, at the lower boundary of cross-sections, we will constrain significantly higher mass macros below
 than did  \cite{PhysRevD.98.115027}. 
 This is simply a result of the increased total exposure obtained  by using the sample of white dwarfs in the MWDD \cite{MWDD}.

\subsubsection{Mass bounds}
Mass constraints can be derived by considering 
the expected number
of macros incident on a sample of white dwarfs or neutron
stars
\begin{equation}\label{mass}
		N_\text{events}=f\dfrac{\rho_\text{DM}}{M_{\chi}}  v_\chi
		\sum_{i=1}^{N_{\text {sample}}}{A_\text{gccs,i} \Delta t_i}\,.
\end{equation}
Here $f$ is the fraction of dark matter comprised of macros
$\rho_{DM}$ is the dark-matter density;
 $M_{\chi}$ is the mass of the macro.
For the $i$-th compact object in the sample:
$A_{gccs,i} = \pi R_{CO,i}^2 (1+v_{esc,i}^2/v_\chi^2)$ is its gravitationally enhanced capture-cross-section 
(for a CO with radius $R_{CO,i}$ 
and surface escape velocity $v_{esc,i}$),
while $v_{\chi}\sim 10^{-3} c$ is the macro velocity
far from the surface;
$\Delta t_i$ is the object's exposure time.

\begin{figure*}
  \includegraphics[width=\textwidth]{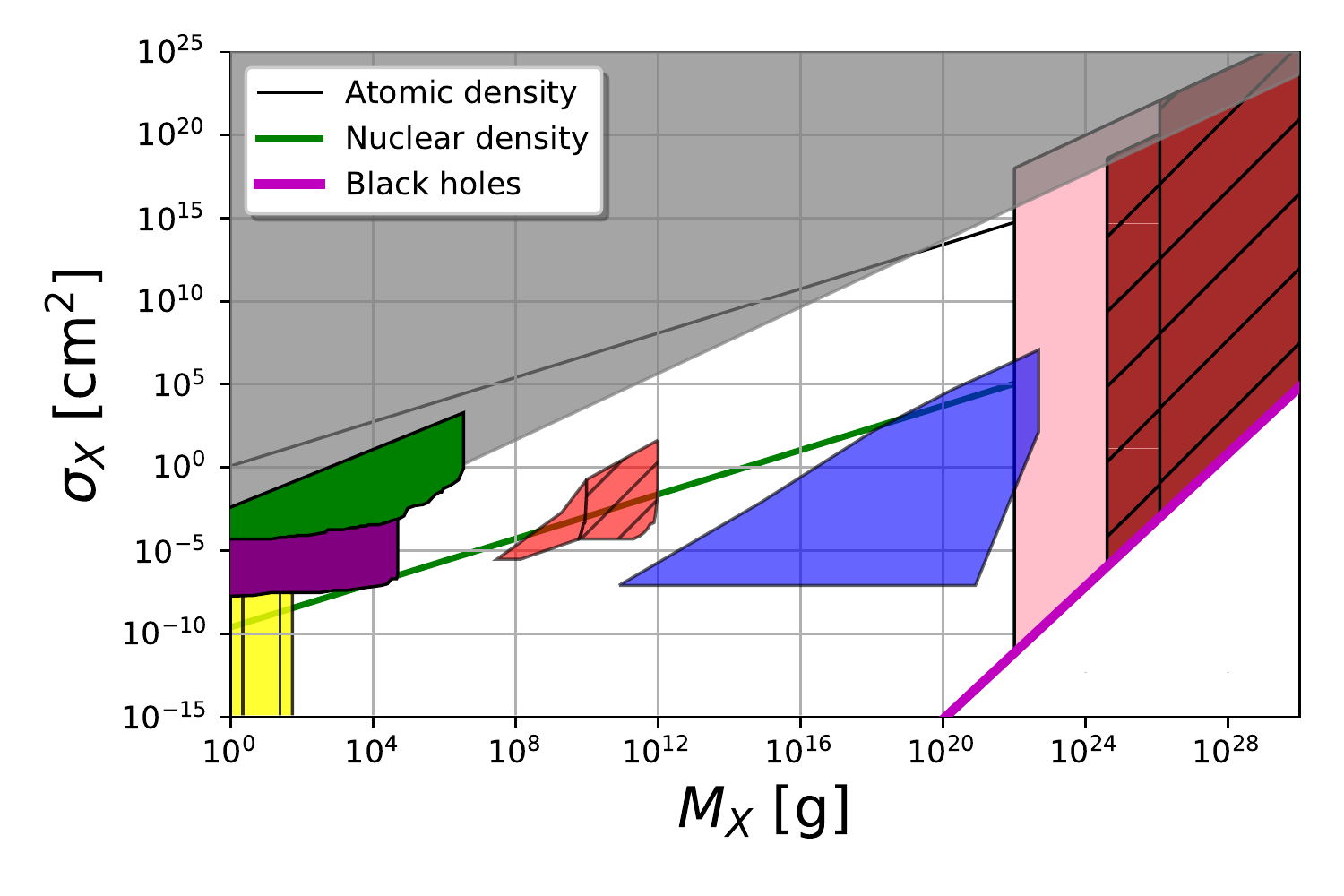}
\caption{Figure 3 of \cite{1610.09680} with the updated constraints discussed in the text. 
Objects within the region in the bottom-right corner should not exist as they would simply be denser than black holes of the same mass. 
The grey region is ruled out from structure formation \cite{1309.7588}; 
the yellow from mica observation 
\cite{DeRujula:1984axn,Price:1988ge}; 
the red from superbursts in neutron stars (this work -- the hatched region representing potential future constraints);
the dark blue from white dwarf becoming supernovae (\cite{PhysRevD.98.115027} as revised in this work); 
the purple from a lack of human injuries or deaths 
\cite{1907.06674}; 
the green from a lack of fast-moving bolides \cite{1908.00557};
the maroon from a lack of microlensing events 
toward the Large Magellanic Cloud and the Galactic center
\cite{Alcock2001,astro-ph/0607207,0912.5297,Griest2013},
and, in  pink, toward M31 \cite{Niikura2019}.
}        
\label{fig:exclusion}
 \end{figure*}

For white dwarfs, 
we use data from the MWDD \cite{MWDD} to place constraints
on more massive macros. 
For each of a sequence of threshold central densities, corresponding to threshold masses,
we apply \eqref{mass} to all white dwarfs in the MWDD 
with known lifetimes 
and masses exceeding the threshold.
The constrained region in Figure \ref{fig:exclusion} 
is the union of the constraints for 
all choice of minimum central density.
The MWDD allows us to push the constrained region 
to higher masses than in \cite{PhysRevD.98.115027} 
or certain cross-section ranges.
Enlarging the MWDD to include more WDs with known lifetimes would extend the range of accessible masses at a given cross section.

For neutron stars, 
the monitoring of X-ray binaries can be used
to constrain lower-mass macros.
Since the low mass X-ray binary 4U 1820-30 exhibited
back-to-back superbursts more than a decade apart \cite{2011ATel}, 
we will use it to place constraints on macros. 
This X-ray binary is located 
approximately 1 kpc from the Galactic center. 
The dark-matter density there 
is expected to be at least $20$ times higher 
than  in the solar neighborhood 
(see e.g. \cite{1012.4515}),
$\rho_{DM} \approx 10^{-17}\,$g m$^{-3}$.

Since macro impacts are approximately a Poisson process, 
the probability $P(n)$ of $n$ macro passages 
through a given neutron star 
over a given exposure time is
\begin{equation}
\label{eq:Poisson}
P(n;N_{events}) = \frac{(N_{events})^n}{n!}e^{-N_{events}}\,,
\end{equation}
where $N_{events}$ 
is the expected number of macro passages through that NS
in that time,
\begin{equation} \label{eventrate2}
		N_\text{events}=
		2\times 10^{11} f \left(\frac{\text{g}}{M_{\chi}}\right)\left(\frac{\delta t}{10\text{yr}}\right)\left(\frac{A_\text{gccs}}{2\times 10^{8} \text{km}^2}\right)\,.
\end{equation}
As expected, 
$A_{gccs} =  \pi (10 { km})^2 (1+v_{esc}^2/v_{\chi}^2)$, 
with $v_{esc} \approx (2/3)c$.
However, we must take care with determining the exposure time, $\delta t$.
After a superburst from 4U 1820-30,
it will take some time $\Delta t$ 
to accrete sufficient column density $y$
from its companion to support another superburst.
$\Delta t = y \pi R_{NS}^2/\dot{M}$,
with 
$\dot{M} \approx 10^{17} g s^{-1}$ 
the accretion rate onto 4U 1820-30.
At a time $T$ after the last superburst, 
the exposure time is 
$\delta t = \max(0,T - \Delta t)$.
Although the time between  superbursts was observed to be
approximately one decade, 
the duty cycle of the instrument that observed these
superbursts, RXTE-ASM, 
is around 40$\%$ \cite{Levine1996}.
Combined with spacecraft maneuvers that were planned to produce 
a highly stochastic pattern of sky coverage, a randomly chosen source was 
scanned typically 5 to 10 times per day \cite{Levine1996},
corresponding to an average time between scans of at most 5 hours.
A typical superburst last around 3 hours \cite{Strohmayer2002}.
This gives an effective duty cycle of $\sim 60\%$.
Thus, there is a non-negligible chance that a 
superburst will be missed.

With this effective duty cycle of $\sim 60\%$, there is less than a $5\%$
probability that we will miss all superbursts in a decade if there 
are at least 4 superbursts during this time. This yields
\begin{equation} \label{dutycycleeventrate}
		N_\text{events}=
		4 \times 10^{10} f \left(\frac{\text{g}}{M_{\chi}}\right)\left(\frac{\delta t}{2.5\text{yr}}\right)\left(\frac{A_\text{gccs}}{2\times 10^{8} \text{km}^2}\right)\,.
\end{equation}

Since no events are observed in this time $T$,
$N_{events} \geq 3$ may be ruled out at 95\% confidence 
since $P(0;3)=0.05$;
this corresponds to
\begin{align}
   M_{\chi} \leq 1 \times 10^{10} f\text{g}\,.
\end{align}

Another superburst constraint could potentially be derived 
by comparing the expected macro-induced rate
for thermonuclear runaway in neutron stars to that observed.
For a population of $N_{NS}$ Milky Way neutron stars 
that are found in compact binaries 
and are accreting from a companion star, 
we expect 
\begin{equation}\label{2ndconst}
	N_{events} = 
	f N_{NS} \frac{\rho_\text{DM}}{M_{\chi}} 
	A_\text{gccs} v_{\chi} t
\end{equation}
macro-induced superbursts in time $t$.
As before, 
if $n$ superbursts have occured, 
where $N_{events}$ were expected,
and $P_{\text {Poisson}}(n;N_{events})\leq0.05$, 
then that value of  $N_{events}$ is ruled out at 
the 95$\%$ level.

Currently, only 15 known neutron stars are known to have experienced a superburst \cite{1702.04899}. 
As the data from observed superburst becomes better, 
we can expect to  probe beyond 
$M_x = 5 \times 10^{10}\,$g. 
For example, the observation of 100 superbursters 
(Figure 12 of reference \cite{Grimm2002} 
indicates there are about 100 Low Mass X-ray Binaries and High Mass X-Ray Binaries respectively), 
each undergoing superbursts no more than twice annually,
would allow $M_x \leq 10^{12}$g to be probed.

It should be noted, that it is possible that some
of the superbursts that are observed are, in fact,
macro-induced!  
However, absent an observable signature 
that distinguishes macro-induced superbursts 
from ordinary superbursts, the best we can do 
is put limits on macro-parameter space 
from the fact that superbursts aren't more common than
observed.

We present our results in Figure 1.
The blue region represents our revised
constraints from white dwarfs.
The red region with no hatching represents constraints 
from  observations \cite{Strohmayer2002}
of 4U 1820-30.
The red hatched region represents constraints
that could eventually be inferred 
from monitoring of neutron stars in X-ray binaries.

\section{Conclusion}
We have applied the analyses 
of \citep{PhysRevD.98.115027} and \cite{Niikura2019} 
to macros 
and identified the regions of cross-section-versus-mass parameter space 
that can be  excluded based on: 
microlensing of stars in M31, 
superbursts in neutron stars, 
and type IA supernova in old white dwarfs. 
Of particular interest, 
parts of the  nuclear-density line in that parameter space
have been ruled out. 
However, there remain three windows for 
nuclear density macros: 
$55\,$g $\lesssim M_{X} \lesssim 10^{3}\,$g,
$5\times 10^4\,$g $\lesssim M_{X} \lesssim 10^{8}\,$g, 
and 
$10^{10}\,$g $\lesssim M_{X} \lesssim 10^{18}\,$g 
A substantial portion of the parameter space above and 
below nuclear density remains unconstrained. 
The atomic-density line is  ruled out,
except for a small window between $10^{20}$g and  $10^{22}$g.

We reiterate that 
certain constraints reported here 
are subject to additional scrutiny 
because it is not certain that the conditions identified
in \cite{Timmes1992} 
are indeed sufficient to initiate thermonuclear runaway, 
i.e. there remains some uncertainty whether in fact
heating a region of size at least $\lambda_{trig}$
to $T \sim$ few $\times 10^9\,$K
necessarily causes type 1A supernovae in white dwarfs
and superbursts in neutron stars. 
We have exercised additional conservatism 
compared to past analyses in deploying that condition
(by taking a larger $\lambda_{trig}$),
however, 
future simulations of the relevant systems
could refine or eliminate the associated constraints.

\acknowledgements
This work was partially supported by Department of
Energy grant de-sc0009946 to the particle astrophysics
theory group at CWRU.
JSS thanks Saurabh Kumar and David Cyncynates for
helpful discussions.
 
 \bibliographystyle{apsrev4-1}
\bibliography{supernovaecites}

\end{document}